# The causal effect of political power on the provision of public education:
# Evidence from a weighted voting system[•]

Erik Lindgren,[♥] Per Pettersson-Lidbom,[♠] and Björn Tyrefors[♦]

## Abstract

In this paper, we estimate the causal effect of political power on the provision of public education. We use data from a historical nondemocratic society with a weighted voting system where eligible voters received votes in proportion to their taxable income and without any limit on the maximum of votes, i.e., the political system used in Swedish local governments during the period 1862-1909. We use a novel identification strategy where we combine two different identification strategies, i.e., a threshold regression analysis and a generalized event-study design, both of which exploit nonlinearities or discontinuities in the effect of political power between two opposing local elites: agricultural landowners and emerging industrialists. The results suggest that school spending is approximately 90-120% higher if the nonagrarian interest controls all of the votes compared to when landowners have more than a majority of votes. Moreover, we find no evidence that the concentration of landownership affected this relationship.

---

[•] This is a revised and extensively rewritten version of the paper "The Role of the Weighted Voting System in Investments in Local Public Education: Evidence from a New Historical Database". This work is financed via ERC consolidator grant no. 616496 (Pettersson-Lidbom).

[♥] Department of Economics, Stockholm University, email: erik.lindgren@ne.su.se
[♠] Department of Economics, Stockholm University and Research Institute of Industrial Economics (IFN), email: pp@ne.su.se
[♦] Research Institute of Industrial Economics (IFN) and Department of Economics, Stockholm University and IFN, email: bjorn.tyrefors@ifn.se.

# 1. Introduction

In this paper, we estimate the causal effect of political power on the provision of public education. We use data from a historical nondemocratic society with a weighted voting system where eligible voters received votes in proportion to their taxable income without any limit on the maximum of votes, i.e., the political system used by Swedish local governments during the period 1862-1909.

This setting offers a number of attractive features for identifying the causal effect of political power on school spending. First, Swedish local governments had a very large degree of fiscal autonomy and were responsible for many important public policies, including primary education.[1] Second, we have annual data on both local school spending and local electoral registers for the universe of approximately 2,400 local governments, which allows us to accurately measure the relationship between political power and school spending. Second, our setting features, similar to many other countries around the world at that time, a sharp conflict of interest between two elites, i.e., agricultural landowners and nonagrarian interests (i.e., emerging industrialists). This allows us to examine important theories of political economy and development. Fourth, and perhaps most importantly, we implement a novel identification strategy where we combine two empirical identification strategies: a discontinuous (unconstrained) threshold regression analysis (e.g., Hansen (1999, 2000) and a generalized event-study design (e.g., Schmidheiny and Siegloch (2020)).

The threshold regression analysis exploits the existence of a discontinuous relationship between the political power of the agricultural landowners and nonagrain interest, at the threshold point of 50% of the vote share of the nonagrarian interests, to identify the effect of political power independently from any smooth (i.e., linear) income effects. The event study also exploits a nonlinearity in the effect of political power for identification, namely, the existence of a discontinuous effect of political power at the time when the treatment occurs, since an event-study design is identical to interaction term analysis. We combine both of these strategies into a generalized event-study design that explicitly takes into account that landowners and nonagrain interests have very different preferences regarding public education.

---

[1] Central governmental grants constituted only approximately 10% of total local government revenues. Instead, the bulk of revenues were raised through a local income tax, and local governments could set the tax rate with complete freedom.

The results from our novel identification strategy suggest that there is no relationship between the share of votes and school spending when landowners have the majority of votes. We interpret this result as saying that when landowners politically control the local government, they only spend the required minimum of school spending as determined by state regulations. In sharp contrast, when nonagrarian interests have more than 50% of the votes, they start spending much more on schools, and the effect increases in proportion with their political power. Indeed, when estimating the cumulative dynamic treatment effect after only 4 years, we find that nonagrarian interests increase school spending by 90-120% in local governments where they control all votes compared to local governments where landowners have the majority of votes.

In the paper, we also provide strong support for our key identification assumptions, i.e., we show that (i) the regression function in our threshold regression analysis is approximately a piecewise linear function of the vote share with one common linear relationship for landowners and another common relationship for nonagrarian interests, (ii) the threshold point is approximately 50% of the vote share, and (iii) the parallel trend assumption is likely to hold for the event-study design for at least 7 years before the treatment occurs.

Our paper contributes to a number of literature categories. At a general level, our paper is related to the seminal literature on the role of institutions in economic development (e.g., Engerman and Sokoloff (1997, 2002), Acemoglu et al. (2001, 2005), and La Porta et al. (1997, 1998)) since we study the consequences of introducing a weighted voting system to human capital-promoting institutions. More specifically, our work is related to the literature on the extension of suffrage (e.g., Acemoglu and Robinson (2000, 2001, 2006), Lizzeri and Persico (2004), Llavador and Oxoby (2005)). We contribute to this literature by studying an externally imposed (i.e., by the state) suffrage reform at the local level in a nondemocratic setting where there exists a stark conflict of interest between two local elites, i.e., the entrenched landed elite and the emerging capitalist elite. Indeed, our study is, to our knowledge, the first to credibly identify that a change in *de jure* political power from landowners to emerging industrialists was a key determinant of the emergence of growth-promoting institutions, such as public schooling, in a nondemocratic setting. This finding is also consistent with our previous work (i.e., Lindgren et al. (2019, 2021)), where we argue that the weighted voting system played a key role in the Swedish growth miracle, i.e., transforming Sweden from one of the poorest countries in Europe in the mid-19th century to one of the richest countries worldwide in the 1960s.[2]

Our paper also contributes to the literature on the importance of human capital formation in the development and growth process (e.g., Barro (2001), Easterlin (1981), Galor, (2005), Glaeser et al. (2004), Goldin and Katz (2008)).[3] This literature has shown that economic inequality can be harmful for economic development and growth (e.g., Galor and Zeira (1993), Engerman and Sokoloff (1997, 2002)). Specifically, our work is related to the work on the relationship between landownership concentration and the provision of public education (e.g., Andersson and Berger (2018), Banerjee and Iyer (2005), Cinnirella and Hornung (2016),

---

[2] In short, our argument is that the introduction of the weighted voting system at the local level shifted political power at town meetings from the entrenched landowners to emerging industrialists. As a result, local government where industrialist started to dominate town meetings begun to invest in both local public education (i.e., this paper) and local railways. In Lindgren et al. (2021), we show that local governments with access to local railways increased their real income with approximately 120% over 30 years, while in Lindgren et al (2019), we show that local government where landowners continued to hold political power, economic development was blocked through the use of labor coercion and factor price manipulation, i.e., using entry barriers and other distortionary policies. In Hinnerich and Pettersson-Lidbom (2014), we also show that landowners could capture town meetings even after democratization.
[3] Ljungberg and Nilsson (2009) analyze the effect human capital using historical Swedish data and time-series methods.

Easterly (2007), Galor et al. (2009), Gallego (2010), Go and Lindert (2010), Ramcharan, (2010)).[4] However, a major empirical challenge for this work is that land inequality is also correlated with many other potential determinants of long-run development, as discussed by Acemoglu et al. (2008). Indeed, Acemoglu et al. (2008) argue that it is particularly challenging to separately identify political inequality from economic inequality since these two factors are often highly correlated and that these issues "have not been addressed by the existing literature."

A key contribution of our paper is that we are able to distinguish between political and economic factors with our two novel identification strategies. In addition, as a further check that our political economy explanation is plausible and not confounded by the issue of inequality in land ownership, we divide the local governments into two groups: those where the concentration of landownership was high and those where it was low. Reassuringly, both our identification strategies produce similar results in these two samples. Thus, we find no evidence that land inequality is an important mechanism in our setting.

The rest of this paper is structured as follows. Section 2 describes the historical background, the weighted voting system, the primary education system, and the data used in the analysis. Section 3 presents the empirical designs and the results. Section 4 analyzes whether the concentration of landownership is relevant for the analysis, while section 5 concludes.

## 2. Background

In section 2.1, we first provide a rather long description of the Swedish setting in the 19[th] century since it provides a background for the existence of stark social conflicts of interest between agricultural landowners and nonagrarian interests (e.g., emerging industrialists). We thereafter describe the weighted voting system (section 2.2), the primary education system (section 2.3), and the data used in the analysis (section 2.4).

---

[4] Andersson and Berger (2018) also use historical data on Swedish local governments to test whether political inequality affects local school spending. However, they only perform a regression control analysis for a single cross-section (1871). Moreover, their results are completely different from ours since they find that local governments governed by large landowners spend more on schools than other types of local government.

**2.1 Sweden in the 19th century**

In the middle of the 19th century, Sweden, a predominantly rural and agricultural-based society, was one of the poorest countries in Europe. For example, almost 80% of its nearly 3.5 million inhabitants worked in the agriculture sector, while less than 10% worked in the industrial and handicraft sectors. In addition, it was not until 1943 that the share of employment in the industrial sector was larger than that in the agricultural sector (Edvinsson 2005).[5] Moreover, much of the early industrialization occurred in rural areas, not in cities. For example, as late as 1901, 64% of the total employment in the industrial sector was based in rural areas. Rural industrialization therefore implied that industrialists and landowners were competing for the same labor pool. Rural areas were also sparsely populated; only 10% of all Swedes lived in the country's 87 small towns in 1850.[6] The health situation was also bad since the life expectancy was only 41 years, and the average infant mortality rate was 15% in 1855 but could be as high as 40-50% in certain rural regions (e.g., Brändström 1984). Thus, overall, in the mid-1800s, Sweden was an economically and socially backward country. Nonetheless, Sweden became one of the richest, healthiest, and most industrialized countries worldwide 100 years later.

Regarding the political system, Sweden was, to a large degree, a semifeudal society in the middle of the 19th century. Specifically, it had a parliament (the Diet) consisting of four estates: nobles, clergy, burghers and landowning farmers.[7] As a result, approximately only 5% of the total Swedish population had some form of political rights in the feudal society. Notably, all types of landowning farmers had political rights, which was in sharp contrast to most other

---

[5] The value added from manufacturing was larger than that from agriculture only after 1920.
[6] There were only 87 towns, and they typically had very small populations, except for the city of Stockholm.
[7] The Parliament Act of 1866 introduced a new system of representation, namely, a bicameral legislature.

European feudal societies. However, the vast majority of Swedish landowners were smallholders. For example, 95% of all landowners had a farm size smaller than 30 hectares, while only 1% had a farm size above 100 hectares in 1870. Thus, the smallholder farms operated 70% of all arable land. In other words, the Swedish agricultural economy was largely characterized by subsistence farming and production for the local market.[8] Nonetheless, both small and large landowners were dependent on a large supply of inexpensive labor for the very short harvest season. Thus, to ensure that landowners had a reliable supply of inexpensive labor, a repressive agricultural system was created by the Swedish feudal elites in the 17th century. As a result, Swedish feudalism could be characterized as a system of labor coercion since all types of landowners made extensive use of the Master and Servant Act, which established that farm servants (e.g., both farm hands and maids) should be contracted for one year at a time and were required to do whatever work the master (e.g., the farmer) deemed necessary.[9]

The institution of farm service was a crucial system for the supply of labor in agriculture. For example, in 1870, farm servants constituted more than 30% of the labor force in the agriculture sector. The Master and Servant Act also allowed for coercive measures, such as corporal punishment and police fetching, when servants did not show up for work. The Master and Servant Act also included strict anti-enticement clauses. Thus, a master had almost complete control over his farm servants (e.g., Eklund 1974, p. 227). It was only on October 24, 1926, that the Master and Servant Act was abolished.

A second important component of the labor-repressive agrarian system was that the common people, e.g., landless agricultural laborers, were required by law to be employed, typically as farm servants; otherwise, they could be imprisoned for life (Eklund 1974, p. 211). In other words, it was forbidden for rural landless people to be unemployed.[10]

---

[8] The typical smallholding was a family farm with permanent hired labor, primarily unmarried farm hands and maids, employed by the year and paid in kind (e.g., free lodging and food), with a very small cash wage (Morell and Myrdal (2011, p. 174)).
[9] Acemoglu and Wolitsky (2011), for example, also argue that European feudalism was primarily a system of labor coercion.

A third component of the labor-repressive agrarian system was that a large share of tenant farmers was required by law to perform corvée labor, i.e., unpaid labor demanded by the landowner. The amount of corvée labor also depended on the size of the tenant farm, with larger farms having more corvée obligations than smaller farms (e.g., Morell 2011). Tenant farmers were typically required to work 3-4 days per week, but in some areas, corvée labor could run as high as 700-800 days of work per year,[11] implying that the household of the tenant farmer either had to be large enough to provide this labor itself or had to subcontract this labor by hiring agricultural laborers and maids. In addition, tenant farmers were required to perform extra work whenever requested by the landowners. This additional work was paid but typically far below the "market" wage. The system of corvée labor was abolished only in 1944, and as late as 1920, nearly 30% of all Swedish farmers were tenant farmers and were therefore basically required to perform corvée labor or extra work at a very low wage.[12]

The labor-repressive element of the Swedish agrarian system was also reinforced by the fact that labor mobility (domestic movements) was severely restricted since Sweden maintained a rigid system of internal passport control until 1860. The poor relief law ("hemortsrätt") further restricted labor mobility among the poorer segments of society until 1956. In addition, freedom of trade was heavily circumscribed until the mid-19th century, when the craft guilds were abolished in 1846 and a more general freedom of trade act for men and unmarried women was introduced in 1864.

To conclude, Sweden had a very labor-repressive agricultural system, and from a European perspective, labor coercion was abolished extremely late. Moreover, Swedish landowners were heavily dependent on inexpensive labor and therefore had strong incentives to block the out-migration of labor and industrialization. As a result, there existed a stark social conflict of interest between landowners and emerging industrialists.

---

[10] This law was abolished in 1885.
[11] See Olsson (2006).
[12] For a description of the Swedish corvée labor system, see Morell (2011).

For example, in the case of promoting public education, landowners were very much against investing in basic education, while emerging industrialists, who were striving for an educated labor force, typically favored investment in public education. There are a number of reasons why Swedish landowners were against basic education as discussed by Larsson and Westberg (2020). Perhaps the most important explanation was that child labor was an essential part of the agricultural work force at both large and small farms in the 19th century.[13]

## 2.2 The weighted voting system

Sweden has a long history of local self-government in rural areas. Historically, there existed approximately 2,400 rural local governments, and their decision-making body was the town meeting, i.e., a direct democratic form of government (e.g., see Hinnerich and Pettersson-Lidbom (2014) and Mellquist (1974)).[14] Thus, eligible voters were gathered at town meetings—at least three times per year—to determine matters of economic importance. The town meeting regulation from 1817 stated that, effectively, only landowners had voting rights at the town meetings (e.g., Sörndal 1941)). Typically, the decisions at the meetings were made by unanimity; sometimes, however, in cases of disagreement, a weighted voting scheme, where voters received votes in relation to their farm size, was used. The size of the farm was measured in terms of the "mantal", which was the basic tax assessment unit of land in use since the 16th century and used until 1861. This type of weighted voting system gave landowners with a large farm only a few more votes than those with a small farm.[15]

---

[13] Sjöberg (1996) also discuss the resistance of farmers to investment in primary education due to their reliance on child labor in the early 20th century.
[14] Sweden also had 87-94 urban towns or "cities" in the latter part of the 19th century. The cities had a different political system from the rural local governments.
[15] For a description of the mantal and how it was used in the local governments, see Lagerroth (1928).

In 1862, the four-estate parliament decided to extend suffrage rights at the local level to other groups, including industrialists, in a new Local Government Act. The rationale behind this new law was the private property principle, i.e., all local taxpayers, including companies, should have voting rights in the local government (e.g., Norrlid (1970)). Moreover, one year later, the four-estate parliament decided that all local taxpayers should receive voting rights in proportion to their taxes paid, without any restrictions on the maximum number of votes.[16] Thus, a single taxpayer could have the majority of votes at town meetings.[17] Interestingly, there was no debate among the four estates in the Diet regarding the extension of suffrage rights to industrialists at the local level in 1862, as discussed by Mellquist (1974, p.52). Similarly, the decision to make votes proportional to taxable income in 1863 was also accepted with unanimity, Mellquist (1974, p. 71). Mellquist (1974) provides an explanation for these nonconflictual decisions to extend and modify the suffrage rights: "because companies were so small and few at that time, it was impossible to foresee the subsequent economic development and industrialization."

Most importantly, the taxable income of a local taxpayer was determined by a uniform nationwide regulation introduced in 1861.[18] Specifically, for landowners, taxable income was set to 3% of the assessed agricultural property value, and they received 2 votes for every 0.10 krona of taxable income. Thus, the fixed rule that determined the votes of landowners was as follows:

(1)    $V_t^L = f(\text{taxable income in period t-1}) = 2 \times (\text{Assessed property value} \times 0.03)/10$

---

[16] For example, one industrial firm (Ljusne Woxna AB, Söderala) had 87,974 votes in 1900; in comparison, the average number of votes per taxpayer was approximately 50.
[17] For example, a single taxpayer had the majority of votes in 54 local governments in 1871 and in 44 local governments in 1892.
[18] Article II in Bevillningsstadgan.

where $V_t^L$ represents landowners' votes in period $t$. For the nonagrarian interests including the industrialists, the taxable income was based income from labor, capital, and the operating profits of firms. The nonagrarian interests received 1 vote for every 0.10 krona of their taxable income. As a result, the fixed rule that determined the votes of industrialists was as follows:

(2)   $V_t^N = g(\text{taxable income in period t-1}) = 2 \times (\text{taxable income}/10)$

where $V_t^N$ represents the votes of the nonagrarian interests in period $t$. Equally importantly, the number of votes each taxpayer would receive, which could be used for voting at town meetings, was updated each year.

Official roll call data from 1871 show that an average of 156 people had voting rights in a typical local government of 1,540 citizens and that 127 received their rights from landownership, while the remaining 29 received voting rights based on other, nonagrarian sources. Importantly, only 9 individuals received voting rights from both landownership and any of the other nonagrarian sources. As a result, there is limited overlap between the two groups of landowners and nonagrarian interest, which makes it plausible to treat them as two separate groups with opposing interests.

The official roll call data also show that the typical group of landowners had an average of 4,312 votes, while the nonagrarian interests had 1,993 votes. Thus, a single voter in the landowning group had on average 34 votes, while a single voter from the nonagrarian group had 69 votes, i.e., twice as many votes. Moreover, the average maximum number of votes a single voter possessed was more than two times larger if the voter belonged to the nonagrarian interests, i.e., 1,179 vs. 572. As a result, it typically required twice as many attendants from the group of landowners for them to constitute a majority of votes at a town meeting. Indeed, nonagrarian interests could often have a large influence on the decision-making process at these meetings since the number of meeting attendances was typically very low, often less than 10 people.

To measure the relative political power of the two groups, we define the *de jure* political power of the nonagrarian interest as their relative share of votes as determined by the annually updated electoral rolls., i.e.,

(3) $$\frac{V_{it}^N}{V_{it}} = \frac{V_{it}^N}{V_{it}^L + V_{it}^N},$$

where $V_{it}^N$ is the number of votes of the nonagrarian interests in local government *i* in year *t* and $V_{it}$ is the total number of votes, i.e., the sum of the nonagrarian votes $V_{it}^N$ and the votes of landowners $V_{it}^L$.

Figures 1 and 2 show the distribution of the share of industrialist votes for 1872 and 1909, respectively. Figure 1 reveals that there were few local governments in 1872 where the industrialists had more than 50% of the votes, i.e., 206 out of 2,336. Figure 3 shows that the number of local governments where industrialists had more than 50% of the votes increased significantly in 1909, i.e., 781 out of 2,354. Thus, the weighted voting system dramatically shifted political power from landowners to industrialists in many local governments during the period 1871 to 1909.

As a result, forward-looking landowners should therefore have had strong incentives to try to block industrialization since they would otherwise become both economic and political losers due to the proportionality between votes and income in the weighted voting system.[19]

Figure 1. Distribution of the vote share of nonagrarian interests in 1872

---

[19] This type of mechanism of political and economic losers is discussed in Acemoglu et al. (2005).

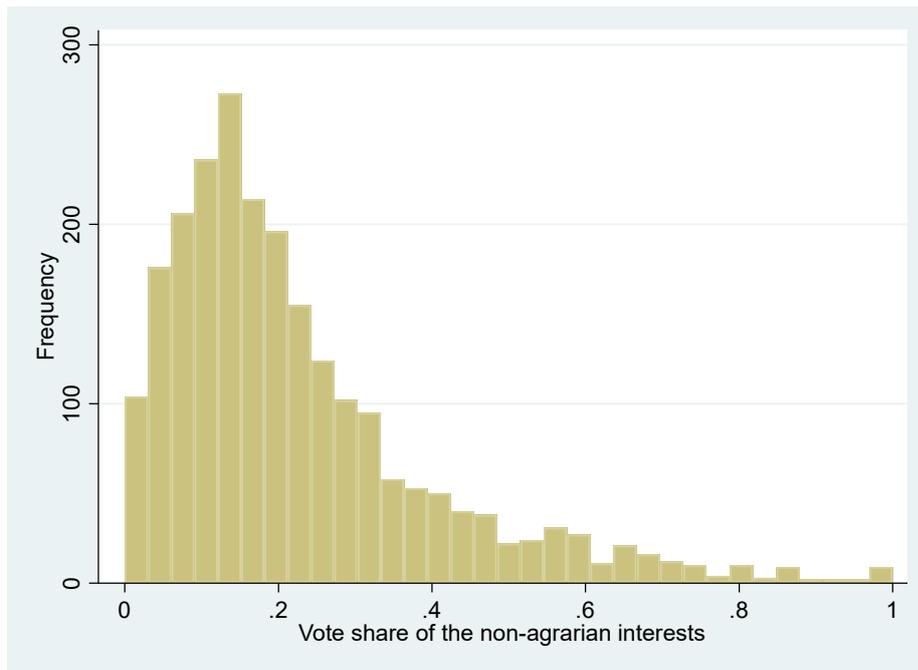

Figure 2. Distribution of the vote share of nonagrarian interests in 1909

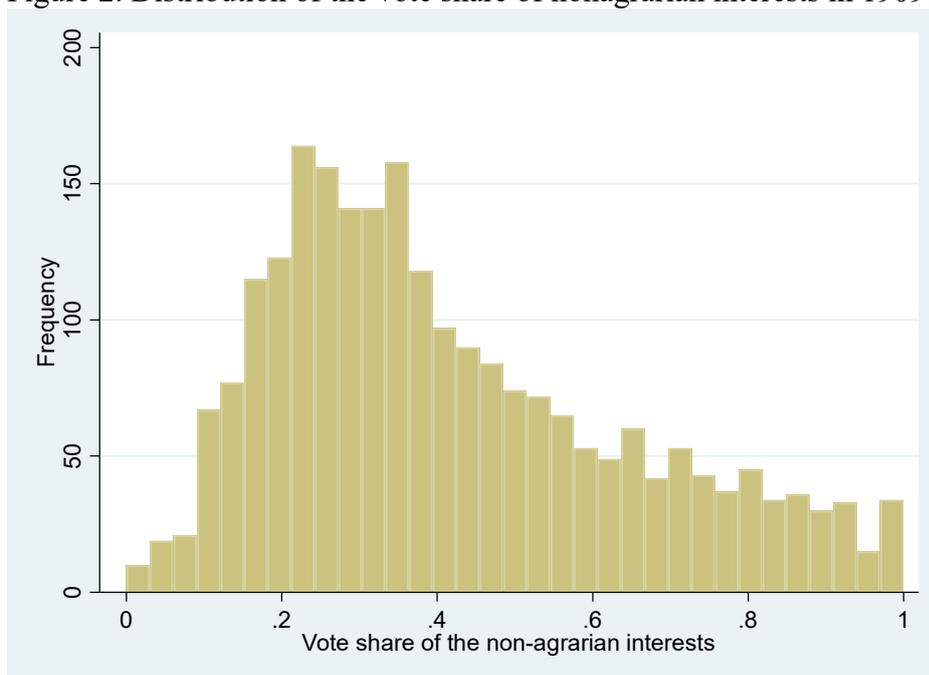

## 2.3 The primary education system[20]

Historically, basic education in Sweden was organized through a system of household instruction, i.e., home schooling, based on the Church Law of 1686, which gave the head of the household the responsibility for educating children and farm workers. Home education was conducted by a household member, and the focus of instruction was on reading and catechetical knowledge. The principal role of the church was to verify and register the level of knowledge via examinations.

---

[20] This section is mainly based on Westberg (2017, 2019).

In 1842, after a long and complicated political process, a school act was initiated that created the main structure for Sweden's primary school system. The school act made it mandatory for each of the 2,308 local governments to organize a school district that operated at least one permanent school within five years.[21] For poor or sparsely populated local governments, the act allowed for an ambulatory school that moved between villages. These ambulatory schools became a main feature of the Swedish school system. The act also stipulated that local governments should establish a school board, which comprised between five and 12 individuals and was led by the parish priest. Moreover, the school board was responsible for the schools within the local government. The board should inspect schools and issue local regulations regarding instruction methods, disciplinary measures, and other school management and organizational issues. The school board was placed under the governance of the town meeting. The town meeting audited its account books and had the final say in issues regarding school expenditures. Thus, the school act placed primary schools in the hands of the local government.

The funding of the primary education system was also largely the responsibility of the local governments since the central government contributed less than 30% to the school revenues for the period 1865 to 1900. The main source of funding during this period was based on a local income tax that local governments could set with complete freedom.

---

[21] The administrative system at the local level is complicated since it consists of at least 5 entities: local governments (borgerliga kommuner), parishes (kyrkliga kommuner), school districts (skoldistrikt), cadastral units (jordebokssocken), and poor relief societies (fattigvårdssamhällen). However, in most cases, these administrative entities are identical; thus, we use the word local governments for all these entities.

Importantly, the school act did not make schooling compulsory for school-aged children. Indeed, it only stated that all local governments should have had established schools by the 1850s. Moreover, the act stipulates that children should start school by age 9, but it did not otherwise regulate the terms of school years or school days. As a result, only approximately 21% of children between 7 and 14 years of age were enrolled in primary schools in 1843. The attendance rate of enrolled children was low, at 34% on average. However, in rural areas, it could be as low as 10% (e.g., Johansson (1972)). Moreover, the average actual school year consisted of only 60 school days in 1843. However, by 1910, the primary education system had improved considerably since the enrollment rate was 75%, the attendance rate was 80%, and the school year had increased to 166 days.

The quality of the early primary education system was also in a very poor state.[22] For example, in 1843, the number of teachers was only 2,800, i.e., one teacher per 130 children (e.g., Richardsson (1992, p. 31)).[23] Moreover, for half of the teachers, their qualifications were not examined. In addition, local landowners (i.e., taxpayers) showed little or no interest in supporting spending on primary schools (Richardsson (1992, p. 32)). As a result, children in primary schools received very limited education in the mid-19th century.

As in many countries, the central government started to take a more active part in providing incentives for the development of primary education in the second half of the 19th century. For example, state school inspectors were installed in 1861, and new school standards were issued in 1878, 1889, 1897, and 1900. The standards were a set of recommendations for the number of school types and curricula. To increase the quality of school buildings, national building plans were also issued. Nevertheless, these new regulations could not alter the decentralized organization of the primary school system. Thus, the power of the Swedish primary school system largely remained in the hands of the local governments in the beginning of the 20$^{th}$ century.

---

[22] Sandberg (1979), however, comes to a completely different conclusion about Sweden's educational system since he argues that Sweden had a "strikingly large stock of human and social capital" in the mid-19th century. His conclusion is based on secondary sources and aggregated data. A completely different picture emerges once one analyzes the primary micro data stored in the National Archives. Moreover, Resnick and Resnick (1977) argue that the literacy criterion used by the Swedish church is flawed and cannot be used to assess the literacy of the population as is done in Sandberg (1979). Indeed, Nilsson (1999) demonstrates with Swedish data that, depending on the measurement used, it is possible to obtain estimates of almost any level of literacy.

[23] In the beginning of the 1880s, the teacher-student ratio had improved to one teacher per 60 children.

It is also noteworthy that spending on primary education was also the largest and most important spending program of Swedish local government since it made up 43% of total spending in 1908. The two other major spending programs were spending on poor relief (23%) and spending on clergy (24%).

## 2.4 Data

Our data come from a newly constructed Swedish historical database that includes extremely rich data on local governments, villages, firms, individuals, etc. The database mainly covers the period 1860-1950. The finished database will include approximately 1 billion observations.[24]

In this paper, we use data from the universe of rural local governments, i.e., more than 2,400 local governments. We have yearly data on the population and spending on primary education covering the period 1874-1909. We also have annual data on the weighted voting system for the period 1881-1909. In addition, we have data on the weighted voting system for 1872.

With these data, we can construct annual data on the real per capita spending on primary schooling. We can also create data on the yearly distribution of political power between landowners and industrialists, as previously discussed.

The maximum number of rural local governments in our data set is 2,405, but we can only use the data from 2,193 governments in our main analysis since sometimes the local government did not report data to Statistics Sweden. Moreover, there were changes in the jurisdictional boundaries of the local governments, which makes it impossible to follow the same local government over time. Table 1 displays the summary statistics of the variables we use in our analysis.

---

[24] Part of the database is financed via ERC consolidator grant no. 616496 (Pettersson-Lidbom).

Table 1. Summary statistics for the period 1881-1909

|  | Mean | St. Dev. | Min. | Max. |
| --- | --- | --- | --- | --- |
| Log real per capita school spending | 0.77 | 0.60 | -5.85 | 4.34 |
| Vote share of the nonagrarian interests | 0.31 | 0.22 | 0 | 1 |
| Number of votes for landowners | 5,710 | 4765 | 0 | 86,887 |
| Number of votes for the nonagrarian interests | 6,075 | 18,654 | 0 | 839,130 |
| Population size | 1,651 | 1655 | 48 | 21,877 |

Notes: These data consist of 2,193 local governments for the period 1881-1909.

## 3. Empirical designs and results

In this section, we describe our two empirical designs for estimating the causal effect of the political power of the nonagrarian interest on school spending, i.e., a discontinuous (unconstrained) threshold regression analysis and an event-study design, and how they can be combined together.

Starting with the unconstrained threshold regression approach, we defined the *de jure* political power of the nonagrarian interest in equation (3) as their relative share of votes, i.e., $\frac{V_{it}^N}{V_{it}}$. Thus, the population regression of interest in our panel data setting can now be written as

(4) $\quad Y_{it} = \alpha_i + \lambda_t + \beta\left(\frac{V_{it}^N}{V_{it}}\right) + \varepsilon_{it},$

where $i$ indexes a local government at time $t$. $Y_{it}$ measures school spending, $\alpha_i$ is a local government-specific effect, and $\lambda_t$ is a time-specific effect. The parameter of interest is $\beta$, which measures the causal effect of political power of the nonagrarian interest (relative to landowners) on school spending.

We cannot, however, directly estimate the causal effect of political power on school spending using equation (4) since the number of votes received is proportional to income in the weighted voting system. Thus, the ratio $\frac{V_{it}^N}{V_{it}}$, will measure both the effect of political power and income on school spending.

To solve this problem of separately identifying the effect of political power on income, we first need to control for both the numerator and denominator of $\frac{V_{it}^N}{V_{it}}$ in equation (4) to avoid the ratio problem discussed by Kronmal (1993), i.e.,

$$(5) \quad Y_{it} = \alpha_i + \lambda_t + \beta \left(\frac{V_{it}^N}{V_{it}}\right) + \pi_1 V_{it}^N + \pi_2 \frac{1}{V_{it}} + \varepsilon_{it}.$$

Equation (5) is now formally equivalent to an interaction effects model between two *continuous* variables, i.e., $V_{it}^N$ and $\frac{1}{V_{it}}$. As a result, equation (5) effectively controls for any linear income effects working through these two variables. Crucially, this form of equation (5) makes it clear that the causal effect of political power on school spending, i.e., the interaction effect *β*, could now be independently identified from any linear income effects if there exists a nonlinearity in the effect of political power on school spending. Indeed, we argue that the effect of political power will change discontinuously when nonagrarian interests have more than 50% of the votes since nonagrarian interests prefer higher school spending than landowners, as discussed previously.

In this case, we can use a panel threshold regression model to identify the effect of political power. Threshold regression models are similar to regression discontinuity models, i.e., a regression kink model, as discussed by Hansen (2017). However, a standard, nonparametric, regression kink design (e.g., Card et al. (2015)) is not well suited for panel data with unit-specific fixed effects, time-varying control variables, models with discontinuous regression functions, and for estimating the causal relationships over the whole range of values of the explanatory variable. We therefore use a discontinuous (unconstrained) threshold regression model and consequently modify equation (5) accordingly as

$$(6) \quad Y_{it} = \alpha_i + \lambda_t + \beta^L \left(\frac{V_{it}^N}{V_{it}}\right) + \beta^N \left(\frac{V_{it}^N}{V_{it}} - 0.5\right) \times 1\left[\frac{V_{it}^N}{V_{it}} > 0.5\right] + \delta 1\left[\frac{V_{it}^N}{V_{it}} > 0.5\right] + \pi_1 V_{it}^N + \pi_2 \frac{1}{V_{it}} + \varepsilon_{it}.$$

where $1\left[\frac{V^N}{V} > 0.5\right]$ is an indicator function taking the value 1 if the nonagrarian interest has more than 50% of the votes and takes the value 0 if landowners have the majority of votes. Thus, $\beta^L$ and $\beta^N$ are the group-specific common slopes measuring the effect of political power on school spending for landowners and nonagrarian interests, respectively.

To probe whether specification (6) is valid, we show covariate-adjusted (i.e., controlling for $\alpha_i$, $\lambda_t$, $V_{it}^N$, and $\frac{1}{V_{it}}$) binned scatterplots between $Y_{it}$ and $\frac{V_{it}^N}{V_{it}}$. Binned scatterplots provide a nonparametric way of visualizing the relationship between two variables that can be used to assess whether there is (i) a sharp change in regression parameters (i.e., $\beta^L$ and $\beta^N$) at a particular threshold point (i.e., 0.5 in our case) and (ii) whether the imposed functional form assumption is plausible, i.e., if the two parameters $\beta^L$ and $\beta^N$ are approximately linear below and above the threshold point. In addition, we will empirically test whether our assumption of a threshold point at 0.5 is correct using the framework developed by Hansen (1999, 2000).

As our dependent variable, we will use the logarithm of per capita school spending. Importantly, we will also control for the logarithm of the population to avoid the ratio problem of having the population in the denominator, as discussed by Kronmal (1993). As a result, our regression parameters, i.e., $\beta^L$ and $\beta^N$, are now only identified from variation in school spending, i.e., the policy outcome of interest, and not from any changes in the population.

Starting with the graphical binned scatter analysis, Figure 3 displays a covariate-adjusted binned scatterplot between the log of per capita spending, $Y_{it}$, and the share of nonagrarian votes, $\frac{V_{it}^N}{V_{it}}$. The number of equally sized bins are set to 100. The dashed vertical line marks the 0.5 threshold point, while the red lines below and above the threshold of 0.5 are linear regression lines based on the binned points. Figure 3 reveals that there is a sharp kink in the conditional mean function (CEF) at approximately 50% of the votes but without any discontinuous regression response ("jump") at the threshold.[25] Figure 3 also reveals that the relationship between school spending and the share of nonagrarian votes is approximately linear both above and below the threshold point of 0.5. Specifically, the estimate from the graphical binned scatter analysis is $\beta^L$ close to zero, while the estimate of $\beta^N$ is very large, approximately 0.45. We interpret these results as saying that when landowners have the majority of votes, they only spend the required minimum of school spending as determined by state regulations. Hence, the flat regression line. In sharp contrast, when the nonagrarian interests have more than 50% of the votes, they start spending much more on schools, and the effect increases in proportion with their political power, i.e., the effect increases by 57% (=exp(0.45)) if the nonagrarian interest has all the votes compared to when landowners have the majority of votes.

As a further check of the graphical binned scatter analysis, we also show a binned scatterplot when the number of equally sized bins is reduced from 100 to 50. Figure 4 displays these results and even shows a tighter relationship between the two regression lines and the scatter points than in Figure 3. Thus, this finding suggests that the slope coefficients $\beta^L$ and $\beta^N$ in equation (6) will be precisely estimated.

---

[25] Thus, the absence of a jump in the CEF at the threshold point means that a regression discontinuity design is not applicable

Indeed, the results from estimating the panel threshold regression model (6), as displayed in Column 1 in Table 2, strongly corroborate the graphical binned scatter analysis. Specifically, we find that the estimate of $\beta^L$ is small, i.e., 0.037, and not statically significant from zero, while the estimate of $\beta^N$ is very large, i.e., 0.414, and highly statically significant from zero. Moreover, testing that $\beta^L = \beta^N$ is also strongly rejected (i.e., $F(2, 2192) = 11.10$, Prob > F = 0.0000).

Figure 3. Nonparametric relationship between political power and log per capita school spending

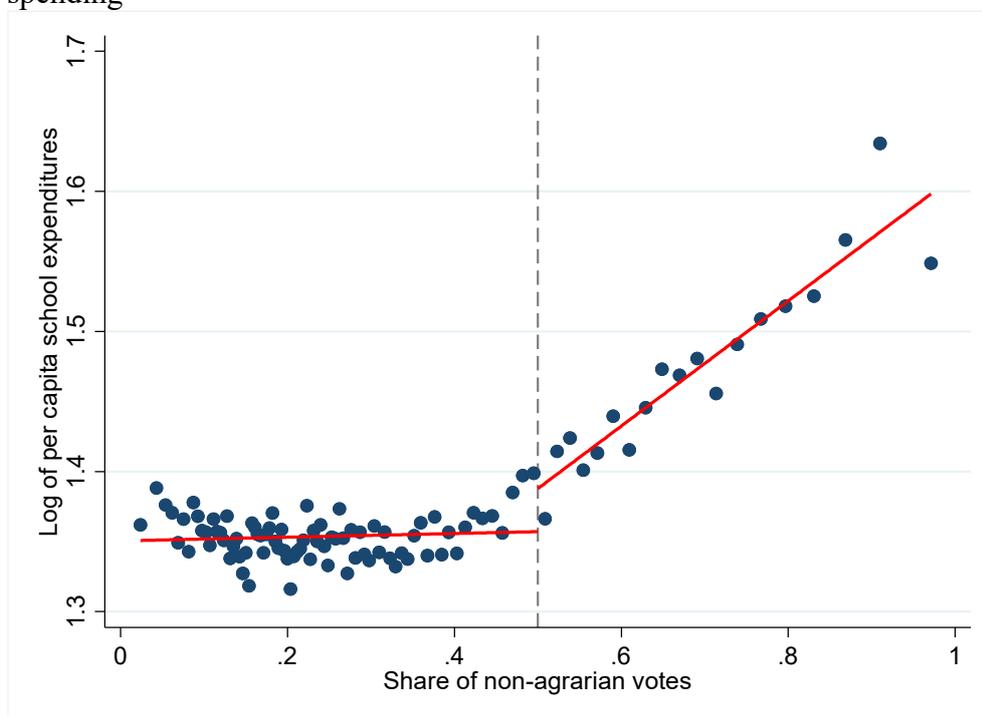

Notes: Binned scatterplot with 100 bins

Figure 4. Nonparametric relationship between political power and log per capita school spending

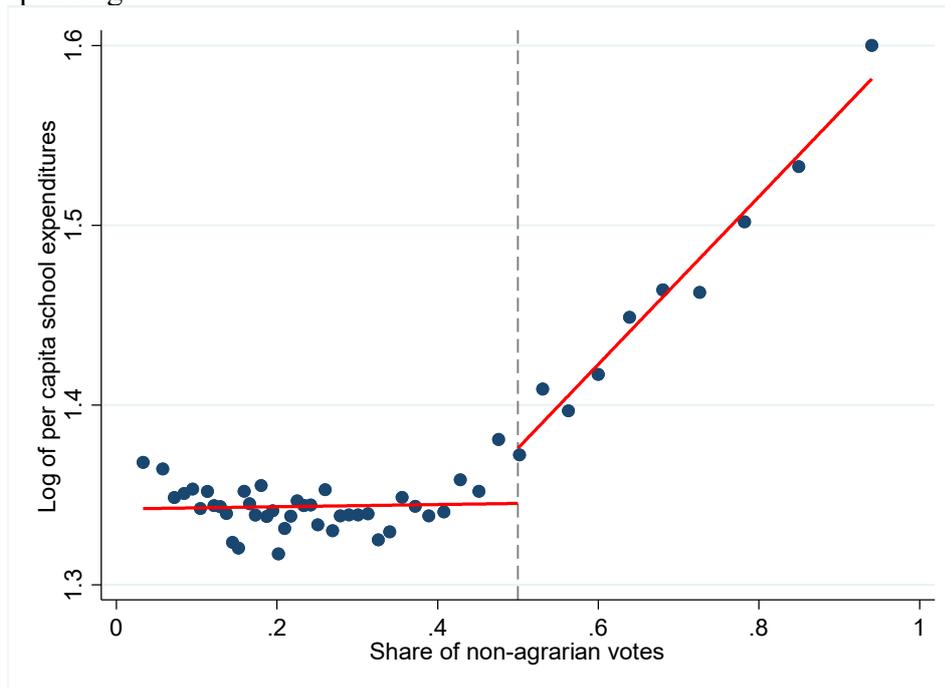

Notes: Binned scatterplot with 50 bins

Thus far, we have assumed that the threshold point is known, i.e., 0.5. However, there are some factors that could potentially make the threshold point deviate from 0.5. One such factor is that some individuals receive votes both from agricultural properties and from nonagrarian sources, as noted above. As a result, this issue will introduce a measurement error in our measure of political power, i.e., the vote share of the nonagrarian interest. Another factor that gives rise to a problem of determining the threshold point *a priori* is that school spending is determined by the meeting participants at regular town meetings and that those participants may not reflect the preferences of the electorate according to the electoral rolls.

To empirically address this issue with a potentially unknown threshold point, we will use the framework developed by Hansen (1999, 2000). We find that the estimate of the threshold point varies between 0.47 and 0.52 depending on the amount of trimming of the data. Reassuringly, our regression results in Column (1) in Table 2 are not much affected by the precise location of the threshold point, as seen from Columns (2) and (3), which show the results from the threshold points 0.47 and 0.52, respectively.

Table 2. Regression results from the discontinuous threshold regression model

|  | Threshold point=0.50 (1) | Threshold point=0.47 (2) | Threshold point =0.52 (3) |
|---|---|---|---|
| Slope coefficient of landowners: $\beta^L$ | 0.037 (0.048) | 0.006 (0.050) | 0.043 (0.047) |
| Slope coefficient of nonagrarian interests: $\beta^N$ | 0.414 (0.097) | 0.422 (0.090) | 0.386 (0.128) |
| Difference in means at the threshold point: $\delta$ | 0.020 (0.012) | 0.023 (0.012) | 0.030 (0.013) |
| Number of local governments | 2,193 | 2,193 | 2,193 |
| Number of observations | 61,702 | 61,702 | 61,702 |

Notes: Table 2 shows the results from estimating equation (6) with different threshold points. Standard errors clustered at the local government level are within parentheses.

We now turn to a discussion of our second identification approach, i.e., an event study design. Importantly, we argued above that nonlinearity in the effect of political power on school spending is required to separately identify the effect of political power on income. Thus, this makes it possible to use an even study design to identify the causal effect of political power since it relies on nonlinearity in the treatment effect at a particular point in time, i.e., the time when the treatment occurs. However, our event study is complicated by the fact that the treatment is not binary but continuous $\frac{V_{it}^N}{V_{it}}$. Nonetheless, it is still possible to use a generalized event study as long as "the treatment effect is proportional to the observed treatment intensity", as discussed by Schmidheiny and Siegloch (2020).

As a result, the generalized event study design and threshold regression analysis rely on different identifying assumptions of the treatment effect. Indeed, the threshold regression analysis assumes that the treatment effect is nonlinear but rules out any dynamic effects, while the event study design considers that the treatment effect is dynamic but restricts it to be linear.

However, we will modify the event study design so that it will be compatible with the threshold regression analysis.

Following Schmidheiny and Siegloch (2020), we estimate the generalized event study using a distributed lag model since they show that the distributed-lag model and an event study design with binned endpoints are numerically identical, leading to the same parameter estimates after correct reparametrization. Specifically, we estimate a distributed lag panel data model with 6 leads and 6 lags,[26] i.e.,

$$(7) \quad Y_{it} = \alpha_i + \lambda_t + \beta_1 \left(\frac{V^N_{it+6}}{V_{it+6}}\right) + \beta_2 \left(\frac{V^N_{it+5}}{V_{it+5}}\right) + \beta_3 \left(\frac{V^N_{it+4}}{V_{it+4}}\right) + \beta_4 \left(\frac{V^N_{it+3}}{V_{it+3}}\right) + \beta_5 \left(\frac{V^N_{it+2}}{V_{it+2}}\right) + \beta_6 \left(\frac{V^N_{it+1}}{V_{it+1}}\right) +$$

$$\beta_7 \left(\frac{V^N_{it}}{V_{it}}\right) + \beta_8 \left(\frac{V^N_{it-1}}{V_{it-1}}\right) + \beta_9 \left(\frac{V^N_{it-2}}{V_{it-2}}\right) + \beta_{10} \left(\frac{V^N_{it-3}}{V_{it-3}}\right) + \beta_{11} \left(\frac{V^N_{it-4}}{V_{it-4}}\right) + \beta_{12} \left(\frac{V^N_{it-5}}{V_{it-5}}\right) + \beta_{13} \left(\frac{V^N_{it-6}}{V_{it-6}}\right) + \varepsilon_{it}.$$

and using clustered robust inference, i.e., the standard errors are clustered at the local government level.

Figure 5 shows the event study plot based on the estimated parameters from equation (7) with normalization to one year prior to the treatment. Most importantly, Figure 5 shows that there are no pretreatment effects (i.e., interaction effects) 7 years before the treatment occurs at time t=0, while there is a sharp (i.e., discontinuous) change in the treatment effect immediately afterwards. The impact effect is 0.15, and the cumulative dynamic treatment effect increases to approximately 0.25-0.30 after 3 years.

However, the assumption of linearity of the dynamic treatment effects is likely to cause the cumulative dynamic treatment effects downwards to be biased downwards if landowners and the nonagrarian interests have very differential responses, as is suggested by the threshold regression analysis.

---

[26] We have 29 years of data (1881-1909), so we are required to make restrictions on the maximum number of leads and lags to include in equation (8) since otherwise we would lose too much data. Without any leas and lags, the number of observations is 61,702, while we only have 35,982 observations when including 6 leads and 6 lags.

Figure 5. Event study regression results for the effect of political power on school spending

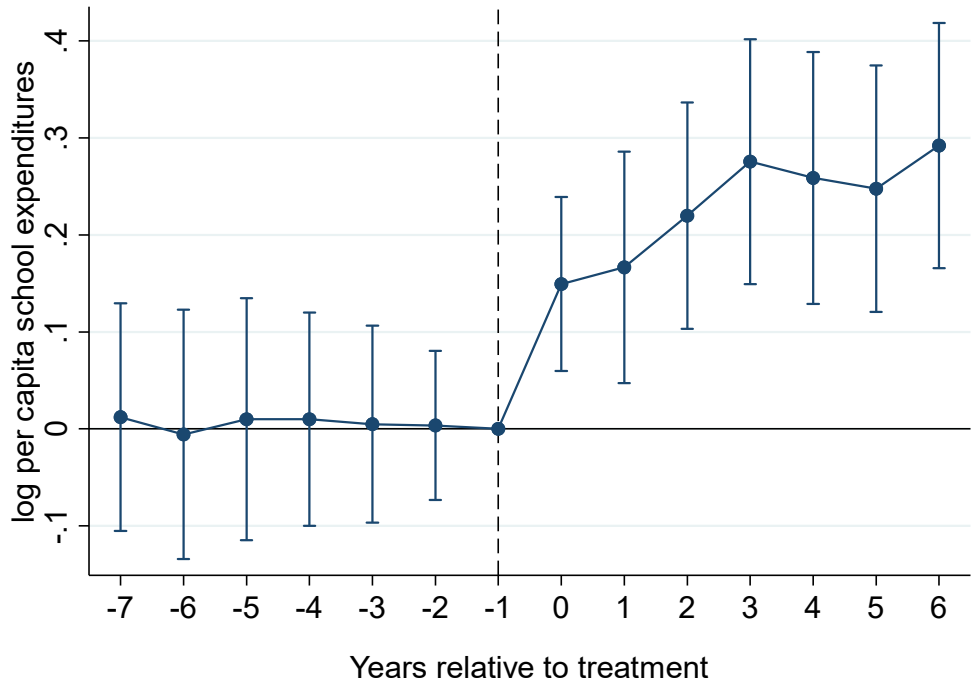

Notes: Point estimates are displayed along with their 95% cluster-robust confidence intervals as described in equation (7). The baseline (omitted) base period is 1 year prior to when the treatment occurs, as indicated by the dashed vertical line in the plot.

It is, however, possible to relax the linearity assumption of the treatment effect in the generalized event study design by combining it with threshold regression analysis. Thus, we can estimate the following distributed lag panel data model:

(8) $Y_{it} = \alpha_i + \lambda_t + \beta_1^L(\frac{V_{it}^N}{V_{it}}) + \beta_1^N(\frac{V_{it}^N}{V_{it}} - 0.5) \times 1[\frac{V_{it}^N}{V_{it}} > 0.5] + \beta_2^L(\frac{V_{it-1}^N}{V_{it-1}}) + \beta_2^N(\frac{V_{it-1}^N}{V_{it-1}} - 0.5) \times 1[\frac{V_{it-1}^N}{V_{it-1}}$

$> 0.5] + \beta_3^L(\frac{V_{it}^N}{V_{it}}) + \beta_3^N(\frac{V_{it-3}^N}{V_{it-3}} - 0.5) \times 1[\frac{V_{it-3}^N}{V_{it-3}} > 0.5] + \beta_4^L(\frac{V_{it}^N}{V_{it}}) + \beta_4^N(\frac{V_{it-4}^N}{V_{it-4}} - 0.5) \times 1[\frac{V_{it-4}^N}{V_{it-4}} > 0.5]$

$+ \beta_5^L(\frac{V_{it}^N}{V_{it}}) + \beta_5^N(\frac{V_{it-5}^N}{V_{it-5}} - 0.5) \times 1[\frac{V_{it-5}^N}{V_{it-5}} > 0.5] + \beta_6^L(\frac{V_{it}^N}{V_{it}}) + \beta_6^N(\frac{V_{it-6}^N}{V_{it-6}} - 0.5) \times 1[\frac{V_{it-6}^N}{V_{it-6}} > 0.5] + \varepsilon_{it},$

where the dynamic treatment effects are allowed to differ between landowners and nonagrarian interests, i.e., $(\beta_1^L, \beta_2^L, \beta_3^L, \beta_4^L, \beta_5^L, \beta_6^L)$ versus $(\beta_1^N, \beta_2^N, \beta_3^N, \beta_4^N, \beta_5^N, \beta_6^N)$. We have also made two other assumptions regarding equation (8), as suggested by our previous analyses. First, we imposed the restriction of no prior treatment effects (i.e., all lead terms were zero), consistent with the evidence from the event study design in Figure 5. We have also tested for a common pretrend by adding lead terms to equation (8), and we cannot reject that they are all zero (these results are reported in the Appendix). Second, there is no discontinuous regression response ("jump") at the threshold point of 50%, consistent with the results from the threshold analysis (i.e., the effect of $1[\frac{V_{it}^N}{V_{it}} > 0.5]$ is close to zero).

Figures 6 and 7 show the results from the effect of political power on school spending using equation (8). These two figures display the cumulative dynamic treatment effects for nonagrarian interests and landowners, respectively, with the normalization that the treatment effect is zero prior to the treatment, i.e., at period *t*-1. Figure 6 shows that there is a large impact effect of 0.46 for the nonagrarian interests, which is similar in magnitude to the effect from the threshold regression analysis reported in Table 2. The cumulative dynamic treatment effect then increases to approximately 0.8 after 4 years. Thus, the log-run effect of political power of the nonagrarian interest on school spending is very large, i.e., 120% (=exp(0.8)). All the cumulative dynamic treatment effects are also statistically significant from zero. In sharp contrast, Figure 7 reveals that there is no effect for landowners since all the cumulative dynamic treatment effects are close to zero and not statistically significant from zero.

Figures 8 and 9 show the corresponding results when we include controls for time-varying confounding effects from income, i.e., $V_{it}^N$ and $\frac{1}{V_{it}}$, and population size in logarithmic form. These time-varying confounders are exactly the same control variables used in the threshold regression approach. Reassuringly, these figures show that cumulative treatment effects are very similar to the previous results, although the long-run dynamic cumulative effect is now somewhat smaller by 90% (=exp(0.65)).

Figure 6. The cumulative effects of political power on school spending for nonagrarian interests

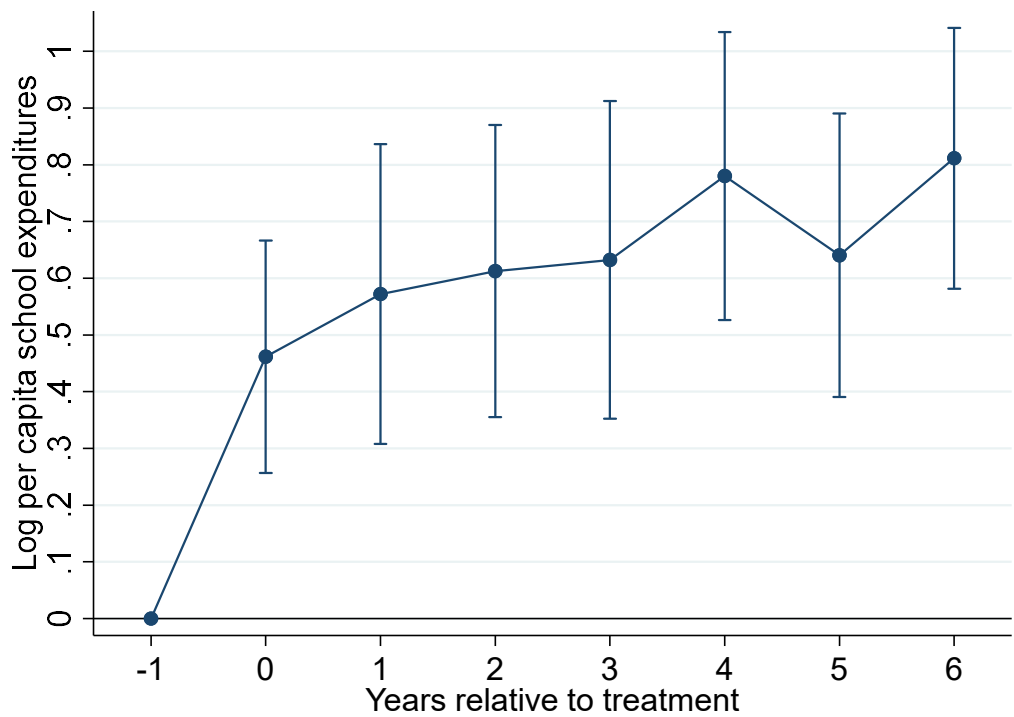

Notes: Point estimates are displayed along with their 95% cluster-robust confidence intervals as described in equation (8). The baseline (omitted) base period is 1 year prior to when the treatment occurs.

Figure 7. The cumulative effects of political power on school spending for landowners

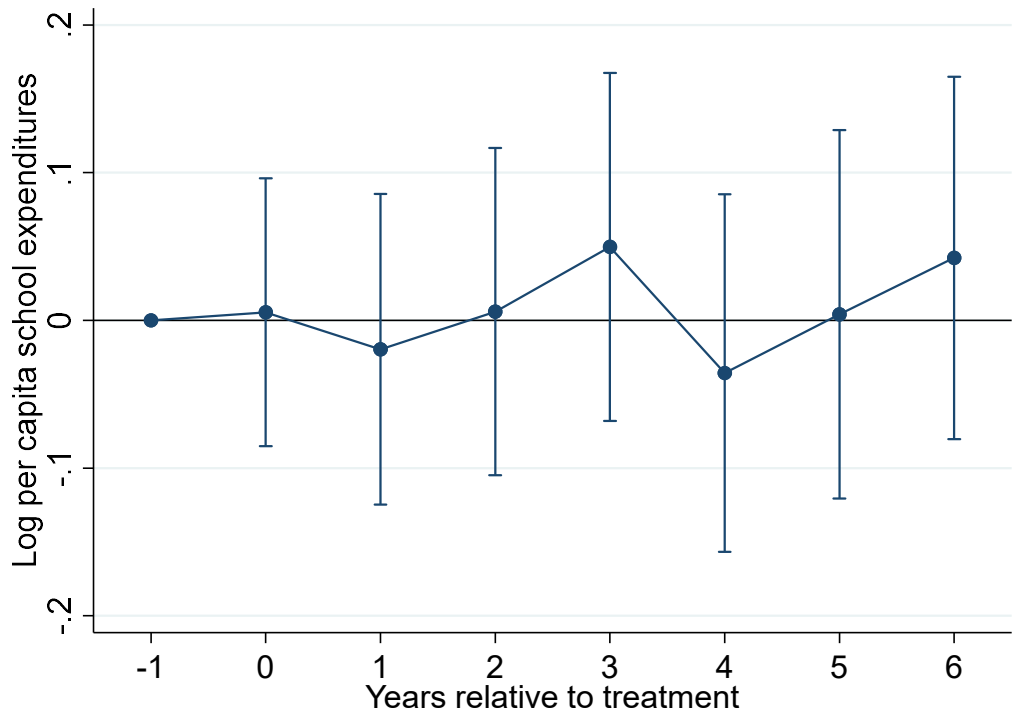

Notes: Point estimates are displayed along with their 95% cluster-robust confidence intervals as described in equation (8). The baseline (omitted) base period is 1 year prior to when the treatment occurs.

Figure 8. The cumulative effects of political power on school spending for nonagrarian interests controlling for time-varying confounders

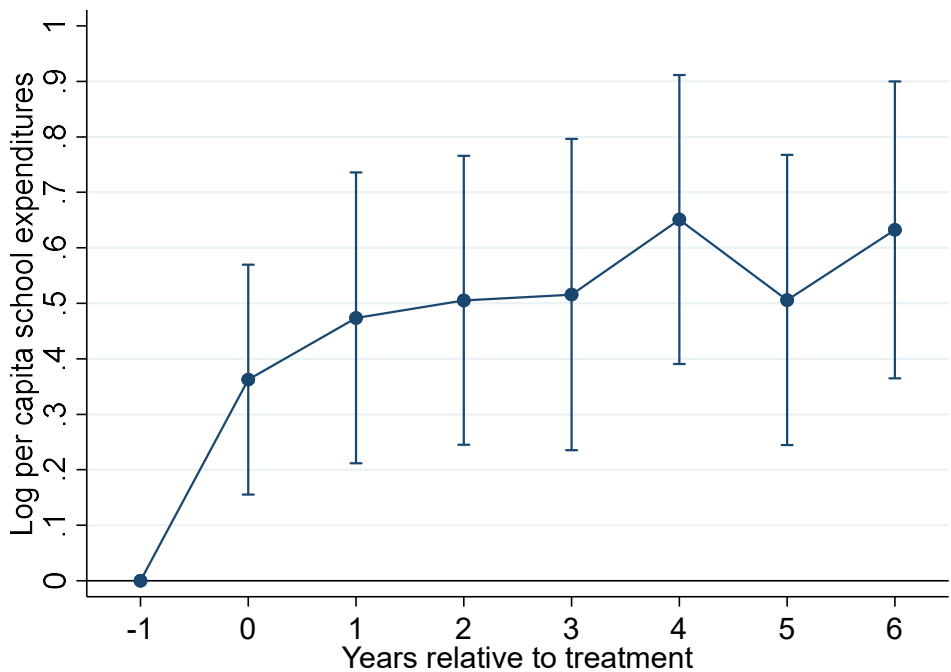

Notes: Point estimates are displayed along with their 95% cluster-robust confidence intervals as described in equation (8). The baseline (omitted) base period is 1 year prior to when the treatment occurs.

Figure 9. The cumulative effects of political power on school spending for landowners controlling for time-varying confounders

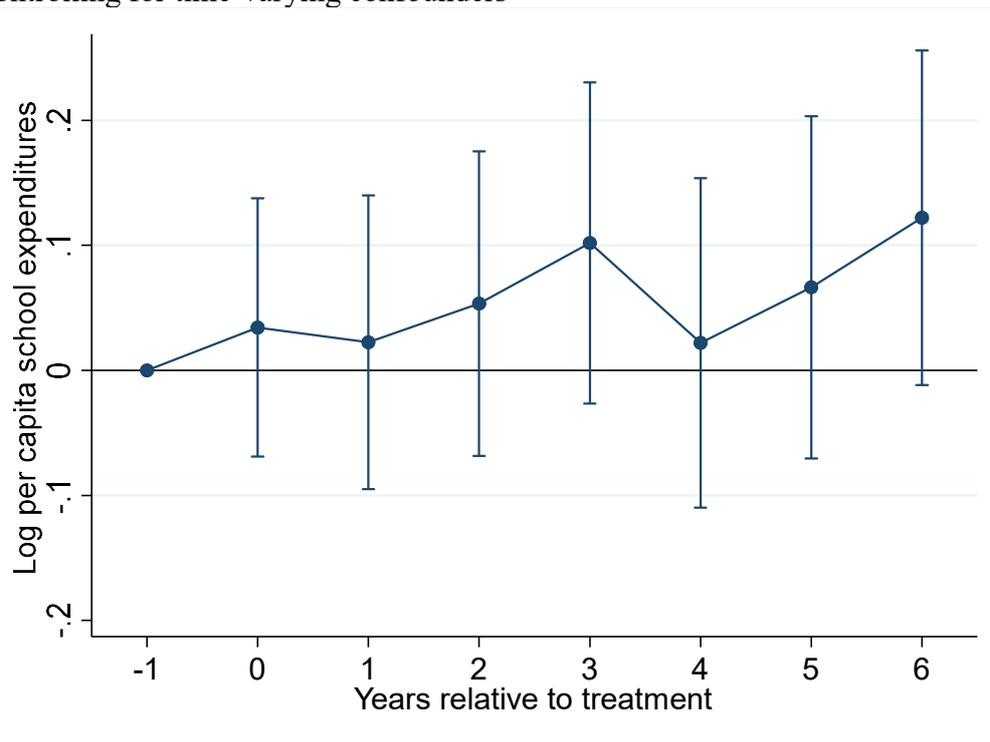

Notes: Point estimates are displayed along with their 95% cluster-robust confidence intervals as described in equation (8). The baseline (omitted) base period is 1 year prior to when the treatment occurs.

# 4. Concentration of landownership

In this section, we analyze whether landownership is of importance for our empirical analysis. As discussed in the introduction, previous research has shown that the concentration of landownership can be an important determinant of school spending (e.g., Andersson and Berger (2018), Banerjee and Iyer (2005), Cinnirella and Hornung (2016), Easterly (2007), Galor et al. (2009), Gallego (2010), Go and Lindert (2010), Ramcharan, (2010)). According to this literature, the political economy channel that links land ownership concentration with the expansion of mass education is that large landowners prevent changes in the supply of schooling by voting against increased school spending.

To test whether this mechanism is of importance for interpreting our results, we perform exactly the same analyses as in our combined identification strategy, i.e., equation (8), but with the only difference that we divide the local governments into two groups: one group where the concentration of large landowners is high and one group where the concentration is low. This classification of landownership concentration is based on the distribution of the assessed agricultural property compiled by Wohlin (1912) for 1865. He classifies a local government where a single landowner has at least 10% of the votes of all landowners as being dominated by large landowners. There are 1,235 local governments that are classified as having a high concentration, while 958 are classified as low.

Figures 10 and 11 show the cumulative dynamic effect of political power on school spending for nonagrarian interests. Figure 10 shows the results for the sample with a high concentration of landownership, while Figure 11 displays the results for the sample with a low concentration. The results from both samples are strikingly similar, i.e., the cumulative dynamic treatment effect is on the order of 0.7-0.9 in both cases. Turning to the corresponding results for landowners, Figures 12 and 13 reveal the exact opposite results, namely, there is no evidence of any cumulative dynamic treatment effects for landowners in any of the two samples.

In summary, we find that the concentration of landownership has little impact on our results. In contrast, our results strongly suggest that both large and small landowners have similar (negative) preferences for school spending. Instead, it is when the nonagrarian interests receive more than 50% of the votes that public education starts to increase dramatically.

Figure 10. The cumulative effects of political power on school spending for nonagrarian interests: Sample with high landownership concentration

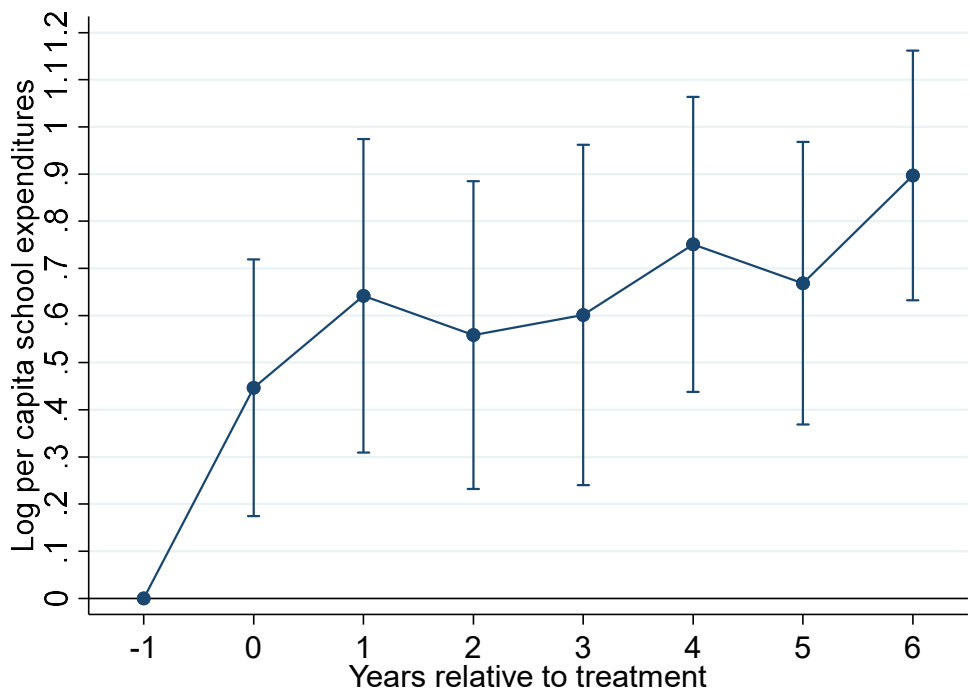

Notes: Point estimates are displayed along with their 95% cluster-robust confidence intervals as described in equation (8). The baseline (omitted) base period is 1 year prior to when the treatment occurs.

Figure 11. The cumulative effects of political power on school spending for nonagrarian interests: Sample with low landownership concentration

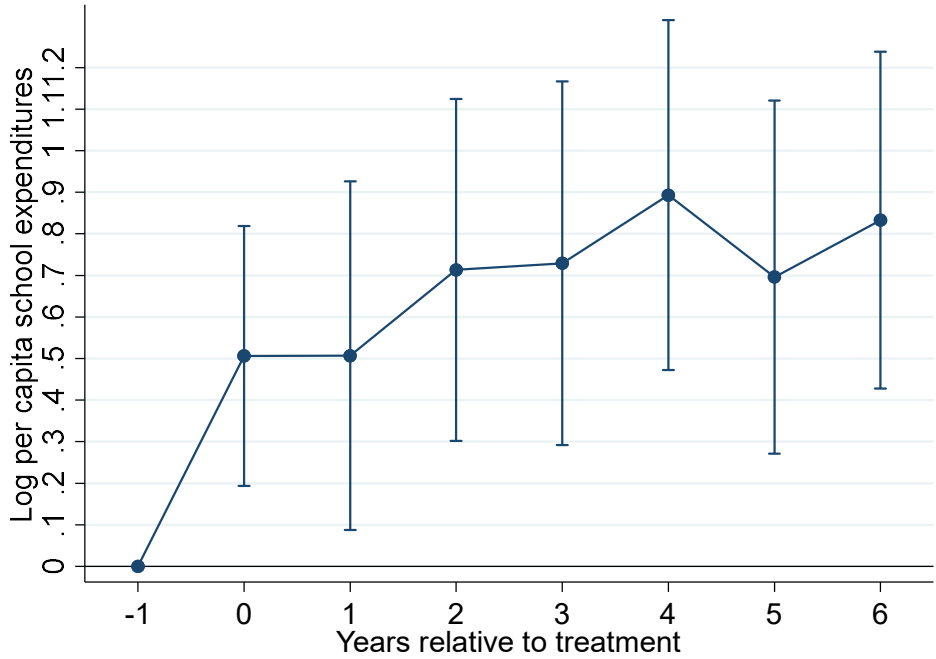

Notes: Point estimates are displayed along with their 95% cluster-robust confidence intervals as described in equation (8). The baseline (omitted) base period is 1 year prior to when the treatment occurs.

Figure 12. The cumulative effects of political power on school spending for landowners: Sample with high landownership concentration

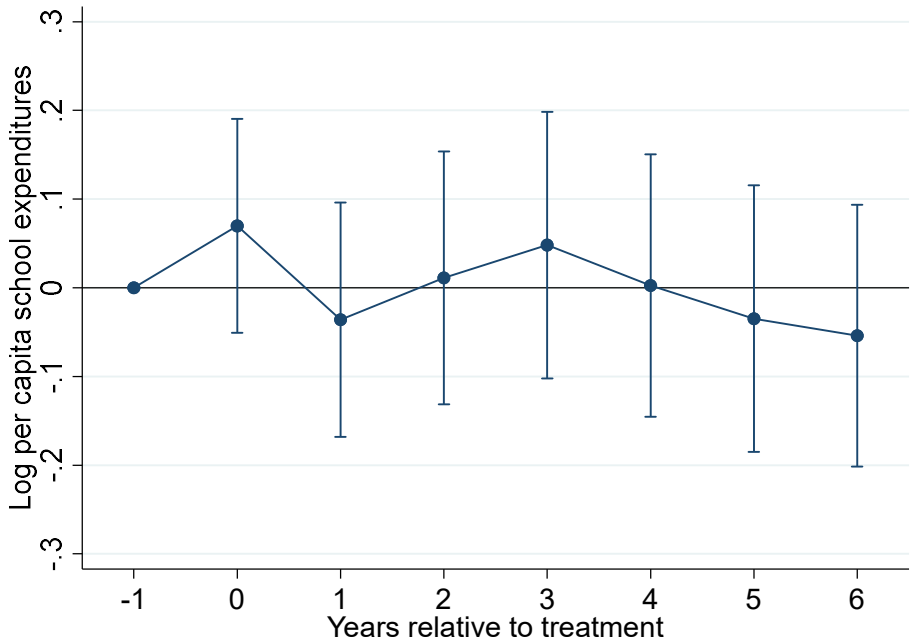

Notes: Point estimates are displayed along with their 95% cluster-robust confidence intervals as described in equation (8). The baseline (omitted) base period is 1 year prior to when the treatment occurs.

Figure 13. The cumulative effects of political power on school spending for landowners: Sample with low landownership concentration

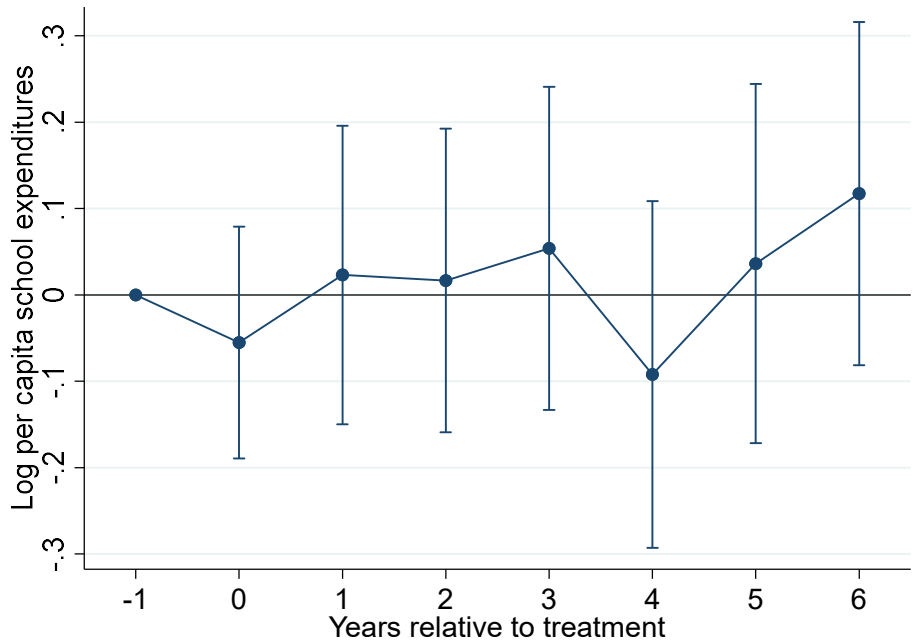

Notes: Point estimates are displayed along with their 95% cluster-robust confidence intervals as described in equation (8). The baseline (omitted) base period is 1 year prior to when the treatment occurs.

## 5. Conclusion

In this paper, we have estimated the effect of political power on public school spending at the local level in Sweden using historical data from a nondemocratic setting with a weighted voting system with no maximum restriction on votes. The Swedish weighted voting system was in place between 1863 and 1909, and it dramatically changed the distribution of political power from landowners to nonagrarian interests (i.e., emerging industrialists)

We use a novel identification strategy where we combine two identification strategies, i.e., a discontinuous threshold regression analysis and a generalized event-study design. We also use a novel data set of approximately 2,200 local governments over the period 1881-1909 that includes annual measures of both school spending and political power of landowners and emerging capitalists.

We show that when nonagrarian interests gain more votes at town meetings, they also start spending much more on primary education than landowners. Indeed, the results suggest that school spending is approximately 90-120% higher if the nonagrarian interest controls all of the votes compared to when landowners have more than a majority of votes. Moreover, we find no evidence that the concentration of landownership affected this relationship.

# **Appendix** (Not for publication)

Here, we report a test for pretrends in equation (8) by including 6 leads to equation (8), i.e.,

$$(8)\ Y_{it} = \alpha_i + \lambda_t + \delta_1 \left(\frac{V^N_{it+6}}{V_{it+6}}\right) + \delta_2 \left(\frac{V^N_{it+5}}{V_{it+5}}\right) + \delta_3 \left(\frac{V^N_{it+4}}{V_{it+}}\right) + \delta_4 \left(\frac{V^N_{it+3}}{V_{it+}}\right) + \delta_5 \left(\frac{V^N_{it+2}}{V_{it+2}}\right) + \delta_6 \left(\frac{V^N_{it+1}}{V_{it+1}}\right) + \beta_1^L \left(\frac{V^N_{it}}{V_{it}}\right) +$$

$$\beta_1^N \left(\frac{V^N_{it}}{V_{it}} - 0.5\right) \times 1\left[\frac{V^N_{it}}{V_{it}} > 0.5\right] + \beta_2^L \left(\frac{V^N_{it-1}}{V_{it-1}}\right) + \beta_2^N \left(\frac{V^N_{it-}}{V_{it-1}} - 0.5\right) \times 1\left[\frac{V^N_{it-1}}{V_{it-1}} > 0.5\right] + \beta_3^L \left(\frac{V^N_{it}}{V_{it}}\right) + \beta_3^N \left(\frac{V^N_{it-3}}{V_{it-}} - 0.5\right) \times 1\left[\frac{V^N_{it-3}}{V_{it-3}} > 0.5\right] + \beta_4^L \left(\frac{V^N_{it}}{V_{it}}\right) + \beta_4^N \left(\frac{V^N_{it-4}}{V_{it-4}} - 0.5\right) \times 1\left[\frac{V^N_{it-4}}{V_{it-4}} > 0.5\right] + \beta_5^L \left(\frac{V^N_{it}}{V_{it}}\right) + \beta_5^N \left(\frac{V^N_{it-5}}{V_{it-}} - 0.5\right) \times 1\left[\frac{V^N_{it-5}}{V_{it-5}} > 0.5\right] + \beta_6^L \left(\frac{V^N_{it}}{V_{it}}\right) + \beta_6^N \left(\frac{V^N_{it-6}}{V_{it-6}} - 0.5\right) \times 1\left[\frac{V^N_{it-6}}{V_{it-6}} > 0.5\right] + \varepsilon_{it},$$

We present the estimates of the pretrends in Figure 1A. All the pretrend estimates are very close to zero and not statistically significant.

Figure 1A. Test of pretrends

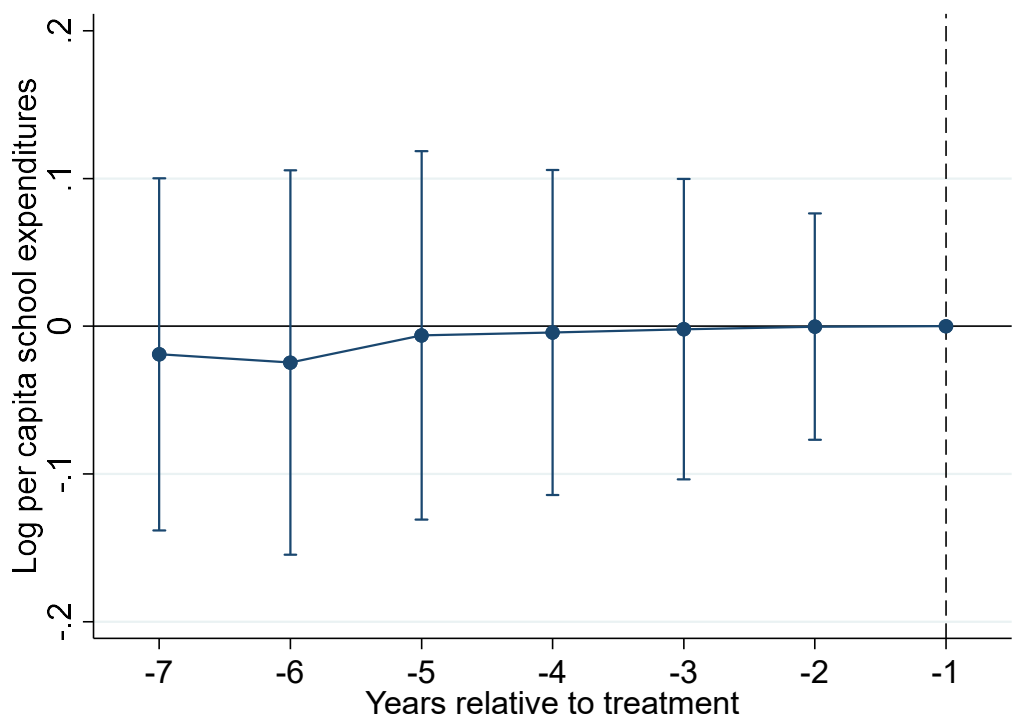